\documentclass[twocolumn
]{aastex631}

\newcommand{\logg}{\log\,g}
\newcommand{\teff}{T_{\rm eff}}
\newcommand{\bz}{\langle B_z \rangle}

\renewcommand{\ion}[2]{#1\,{\sc #2}}

\received{\today}
\revised{\today}
\accepted{\today}

\submitjournal{ApJL}

\shorttitle{The magnetic field of $\upsilon$\,Sgr}
\shortauthors{Hubrig et al.}
\graphicspath{{./}{figures/}}

\begin{document}

\title{The magnetic field of the stripped primary in the $\upsilon$\,Sgr system,
a member of the rare class of hydrogen-deficient binaries}

\author[0000-0003-0153-359X]{Swetlana Hubrig}
\affiliation{Leibniz-Institut f\"ur Astrophysik Potsdam (AIP), An der Sternwarte~16, 14482~Potsdam, Germany}

\author[0000-0003-3572-9611]{Silva~P.~J\"arvinen}
\affiliation{Leibniz-Institut f\"ur Astrophysik Potsdam (AIP), An der Sternwarte~16, 14482~Potsdam, Germany}

\author[0000-0002-0551-046X]{Ilya Ilyin}
\affiliation{Leibniz-Institut f\"ur Astrophysik Potsdam (AIP), An der Sternwarte~16, 14482~Potsdam, Germany}

\author[0000-0002-5379-1286]{Markus Sch\"oller}
\affiliation{European Southern Observatory, Karl-Schwarzschild-Str.~2, 85748 Garching, Germany}

\begin{abstract}

We present the results of high-resolution spectropolarimetric 
observations of the optically dominant component in the rare hydrogen-deficient
binary system $\upsilon$\,Sgr. Only a small number of such systems
in a very late phase of helium shell burning are currently known.
The mass transfer from the donor star in binary systems usually leads to the stripping of its 
hydrogen envelope. Consequently, since the mass of the secondary increases, it appears rejuvenated. 
Using a few ESO FORS\,1 low-resolution spectropolarimetric observations
of this system, Hubrig et al.\ announced in 2009 the presence of a magnetic field of the order of $-$70 - $-$80\,G.
Here we report on more recent high-resolution ESO HARPS spectropolarimetric observations showing that the primary
in $\upsilon$\,Sgr is a spectrum variable star and possesses a weak magnetic field of the order of a 
few tens of Gauss.
The detection of a magnetic field in this rare hydrogen-deficient binary is of particular interest, as
such systems are frequently discussed as probable 
progenitors of core-collapse supernovae and gravitational-wave sources.
Future magnetic studies of such systems will be worthwhile to gain deeper 
insights into the role of magnetic fields in the evolution of massive stars in binary systems.
\end{abstract}


\section{Introduction} \label{sec:intro}

The binary system $\upsilon$\,Sgr (=HD\,181615) with an orbital period of 137.9343\,d was frequently classified in 
the past as a Be star due to the presence of a strong variable emission in the H$\alpha$ line
\citep[e.g.][]{Dudley1993,Koubsky2006}.
\citet{Koubsky2006} suggested that this system is probably 
observed in the initial rapid phase of mass exchange between the components.
The primary, the hydrogen-deficient star dominating the optical spectrum, is less massive with a most 
probable spectral type B5II--B8II \citep{Bonneau2011}, while the other, less visible 
component, is more massive by a factor of 1.57 \citep{Koubsky2006}. 
Recently reported optical interferometric 
observations  of this binary system by \citet{Hutter2021} provided the first direct, visual detection of the hot secondary star,
which is by $3.59\pm0.19$  magnitudes fainter than the primary. \citet{Hutter2021} assigned for the secondary 
star spectral type B5 under the assumption that it is a main-sequence star.
The stellar 
parameters of the primary in the $\upsilon$\,Sgr system, $\teff=12\,300\pm200$\,K and $\logg=2.5\pm0.5$, were reported by
\citet{Kipper2012}.
Apart from the strong deficiency of hydrogen, the primary shows a very 
peculiar chemical composition, with an overabundance of nitrogen and the s-process elements Y, Zr, and Ba.  
The optical spectrum is extremely rich, exhibiting numerous lines
of the ionized metals Ti, Cr, and Fe \citep[e.g.][see also Fig.~7 in \citealt{Hubrig2009}]{Dudley1993,Kipper2012}.
This richness is due to the low hydrogen abundance and consequently low continuum opacity
\citep[e.g.][]{Dudley1993,Kipper2012}.
Interferometric observations in the mid-IR revealed the presence of dust confined to a dense 
circumbinary optically thick disk with an inner rim of about 20\,mas \citep{Netol2009}.
Due to the intermediate 
orbital inclination of the system of about 50$^{\circ}$, it was suggested that the observed spectrum is a combination of 
the disk rim and disk face \citep{Netol2009}.

Hydrogen-deficient binaries (HdBs) are frequently discussed as probable progenitors of core-collapse supernovae 
\citep[e.g.][]{Dudley1993} and possibly gravitational-wave sources \citep[e.g.][]{Laplace2020}.
Recent scenarios for Type~Ib and IIb supernovae consider binary evolution, where the primary star expands as it evolves until
it fills its Roche lobe and then starts mass transfer due to Roche lobe overflow \citep[e.g.][]{Yoon2017}.
The mass transfer from the donor star usually leads to the stripping of the hydrogen envelope.
While the secondary's mass increases due to mass accretion,
this component appears rejuvenated. HdBs are very rare and only a small number of such systems
in a very late phase of helium shell burning are currently known.
\citet{Schoo2018} reported on a set of five such  systems with $\upsilon$\,Sgr the most massive one in this category. 
According to the authors, the donor star in $\upsilon$\,Sgr fills its Roche lobe again during a late
helium shell burning phase, in which it cools and expands. Interferometric observations presented 
by \citet{Bonneau2011} suggest a radius of $18-49\,R_{\odot}$ for the primary, which is much smaller than the $80-130$\,$R_{\odot}$ estimated 
by \citet{Koubsky2006} for the  Roche-lobe radius. Also the more recent interferometric observations by \citet{Hutter2021}
suggest a smaller radius of about 24\,$R_{\odot}$.
According to \citet{Dudley1993}, the mass of the secondary component is 6.2\,$M_{\odot}$ and the radius is about 4\,$R_{\odot}$.
The appearance of the frequently discussed strong variable H$\alpha$ emission in the system $\upsilon$\,Sgr 
is probably due to the material originating from the primary during a
last phase of intense mass transfer and subsequent storage of hydrogen in a stable accretion disk orbiting
around the secondary \citep{Bonneau2011}.

The possible presence of magnetic fields in the complex HdB systems was never considered as one of the critical 
ingredients in previous theoretical and observational studies.
Therefore, this important terrain remained  essentially unexplored.
Based on a few low-resolution spectropolarimetric observations ($R\approx2000-4000$) obtained in 2005 and 2007 at the 
European Southern Observatory with the 
multi-mode instrument FORS\,1 \citep{Appenzeller1998} installed at the 8\,m Kueyen telescope,
\citet{Hubrig2007,Hubrig2009} suggested 
that $\upsilon$\,Sgr is probably a magnetic variable star.
The highest longitudinal magnetic field strengths of
$\langle B_z \rangle=-78\pm8$~G and $\langle B_z \rangle=-73\pm9$~G were measured 
in the FORS\,1 low-resolution polarimetric spectra 
recorded respectively on 2007 August 21 and 31 using for the measurements  
the whole spectrum, including all available absorption lines.

An unsuccessful attempt to confirm the magnetic nature of $\upsilon$\,Sgr was reported by \citet{Silvester2009}, who
obtained three high-resolution ($R\approx65\,000$) observations with the Narval 
spectropolarimeter installed  at the Telescope Bernard Lyot on  Pic du Midi. 
The polarimetric spectra recorded on 2008 June 18--21 at orbital phases corresponding to 
conjunction (where the secondary was behind the primary) were analysed using the least-squares deconvolution
multiline analysis method 
\citep[LSD;][]{Donati1997}, assuming that the lines used in this method have an identical 
shape and that the resulting profile is scaled according to the line strength and 
the sensitivity to the magnetic field. Unfortunately, to create the line mask for the field measurements
using the Vienna Atomic Line Database 
\citep[VALD3; e.g.,][]{Kupka2011}, the authors 
assumed for $\upsilon$\,Sgr a wrong spectral type B2Vpe with 
 $\teff=23\,000$\,K instead of spectral type B5II--B8II with $\teff=12\,300$\,K, and
in spite of the fact that the optical spectra of $\upsilon$\,Sgr contain numerous spectral lines 
of ionized metals usually formed in much cooler atmospheres. 

Here, we report the results of more recent high-resolution ($R\approx115\,000$) spectropolarimetric observations of
this unusual system, acquired using the High Accuracy Radial 
velocity Planet Searcher (HARPS\-pol) fed by the ESO 3.6-m telescope.   

\section{Observations and results}
\label{sect:obs}

\begin{table}
\centering
\caption{
  Logbook of the observations. The columns give the barycentric Julian date (BJD), the orbital phase 
  $\varphi$ calculated using 
  $P_{\mathrm{orb}}=137.9343$\,d and the phase origin  HJD$_{0}=2433018.13$ reported by \citet{Koubsky2006}, 
and the signal-to-noise ratio ($S/N$) measured around 4700\,\AA{}. 
}
\label{tab:obs}
\begin{tabular}{ccr}
\hline
BJD & $\varphi$ & $S/N$ \\ 
\hline
2457093.92148 & 0.545 & 1349 \\
2457094.88906 & 0.552 & 2449 \\
2457177.71636 & 0.153 & 958 \\
2457178.92998 & 0.162 & 2192 \\
2457179.94822 & 0.169 & 1200 \\
\hline
\end{tabular}
\end{table}

The observations were carried out in the framework of the ESO large programme ``Magnetic fields in OB stars''
in 2015 using HARPS\-pol in visitor mode.
The spectropolarimetric observations with this instrument
usually consist of four subexposures observed at different positions of
the quarter-wave retarder plate. The HARPS\-pol spectra cover the spectral range $3780-6910$\,\AA{},
with a small gap between 5259 and 5337\,\AA{}.
The data reduction was
performed using the HARPS\-pol reduction software available
on La Silla and the normalization of the spectra to the continuum level is described in detail by \citet{Hubrig2013}.
The barycentric Julian date (BJD) for the middle of the
exposure, the corresponding orbital
phase, and the signal-to-noise ratio ($S/N$) of the
HARPS\-pol Stokes~$I$ spectra measured at about 4700\,\AA{} are presented in Table~\ref{tab:obs}.

A detailed description of the assessment of the longitudinal magnetic field measurements using 
HARPS\-pol is presented in our previous 
papers \citep[e.g.][]{Hubrig2018,Jarvinen2020}. 
Similar to our previous studies, to increase the $S/N$,
we employed the least squares deconvolution 
technique.
In the spectrum of $\upsilon$\,Sgr,
numerous lines with a low excitation energy of the lower level 
show emission wings, indicating the presence of
circumstellar gas around the star.
Thus, special care was taken to populate our line masks with blend-free lines or lines that are to a lesser extent 
contaminated by the emission lines.
Using VALD3, 
we  created five line masks for five elements, hydrogen (excluding from the mask the strongly variable
emission H$\alpha$ line), \ion{He}{i}, and the iron-peak elements 
\ion{Ti}{ii}, \ion{Cr}{ii}, and \ion{Fe}{ii}, all based on the stellar 
parameters of $\upsilon$\,Sgr, $\teff=12\,300\pm200$\,K and $\logg=2.5\pm0.5$ \citep{Kipper2012}.

\subsection{Line profile variability}
\label{sect:var}

\begin{figure}
\centering 
\includegraphics[width=0.45\textwidth]{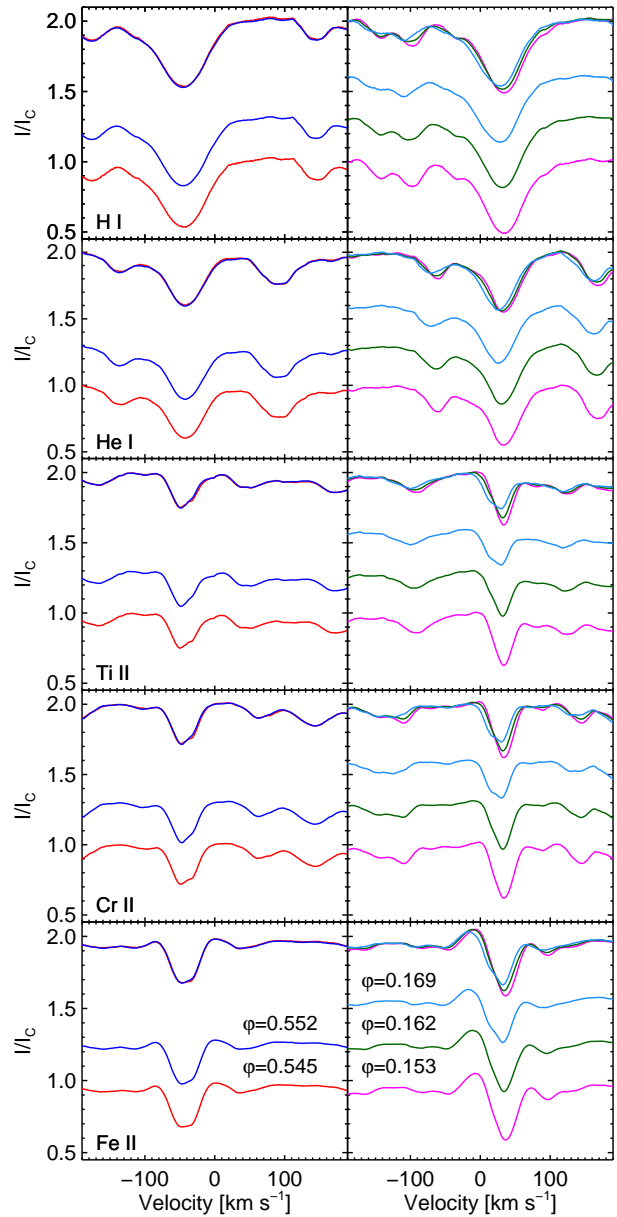}
\caption{
LSD Stokes~$I$ spectra calculated for different line masks using observations acquired 
on five orbital phases.
The spectra are grouped in the left and right panels according to their closest neighbors in phase
and offset in vertical direction for better visibility.
On the top of each panel we present the overplotted spectra to highlight the spectral variability
observed already at nearby orbital phases.
}
   \label{fig:var_all}
\end{figure}

As a first step, to get a better insight into the complex spectrum composition and the spectral variability of 
$\upsilon$\,Sgr, we investigated the
line profile variability using the LSD Stokes~$I$ line profiles 
based on the different line masks. The LSD Stokes~$I$ spectra calculated for each of the five elements and sorted by similar 
orbital phases are presented in Fig.~\ref{fig:var_all}.
The comparison of the line profiles calculated for each phase reveals that the primary is a spectrum variable star
with small changes appearing already at the time scale of one night. The LSD Stokes~$I$ profiles calculated for 
the metal lines, especially those calculated for Ti and Cr, appear asymmetric and show a structure in the line cores.
If we assume that the changes in the line profiles of the metal lines are related to surface chemical patches,
then the rotation period of the primary component should not be too long, of the order of two or three weeks.
Using for the projected rotation velocity the value
$v \sin i=33\pm5$\,km\,s$^{-1}$ given  by \citet{Silvester2009},
the stellar radius of 24\,$R_{\odot}$ determined by \citet{Hutter2021}, and the 
disk inclination of about 50$^{\circ}$ obtained by \citet{Netol2009}, the rotation period is about 28\,d.
No Transiting Exoplanet Survey Satellite (TESS; \citealt{Ricker2015}) or Kepler \citep{Borucki2010} data have been obtained for $\upsilon$\,Sgr so far.
However, TESS should observe this target in Sector~54 in the middle of 2022.

According to \citet{Dudley1990}, the fainter, more massive component, is best visible in the far-UV spectra 
recorded with the IUE satellite. These spectra have been used by these authors to determine the orbit for the
secondary component. To be sure that the secondary component is indeed invisible in our optical spectra, we tried
to isolate the contribution of the main sequence hotter secondary with 6.2\,$M_{\odot}$ by calculating
the LSD Stokes~$I$ profiles using the line mask corresponding to a higher effective temperature,  $\teff=17\,000$\,K
assuming the spectral type B4\,V.
However, no trace of this hotter second component was detected in our LSD spectra.

It is not clear whether a number of smaller absorption features observed bluewards and redwards from the absorption profile
of the primary in Fig.~\ref{fig:var_all} could be assigned
to the accumulated matter in the surrounding envelope where it can behave as a pseudoatmosphere with lines formed in the 
circumstellar environment.  The appearance of similar absorption features was also mentioned by \citet{Silvester2009},
who suggested that the distortion of the continuum could be due to the smaller number of lines in the mask.
However, our tests with line masks containing different numbers of lines confirm the persistence of the features, 
suggesting that they are probably related to the system's circumstellar environment.
The variety of absorption features observed in the LSD Stokes~$I$ spectra 
could also indicate
that different lines trace the circumstellar matter differently, with lines belonging to certain 
elements formed in a region closer to the primary's stellar surface while the matter traced by other lines is farther away 
from the stellar surface.
In any case, future studies are necessary to better understand the structure of this system  including the location and the 
morphology of the line-forming regions that trace different components of the system.

\begin{figure*}
\centering 
\includegraphics[width=0.995\textwidth]{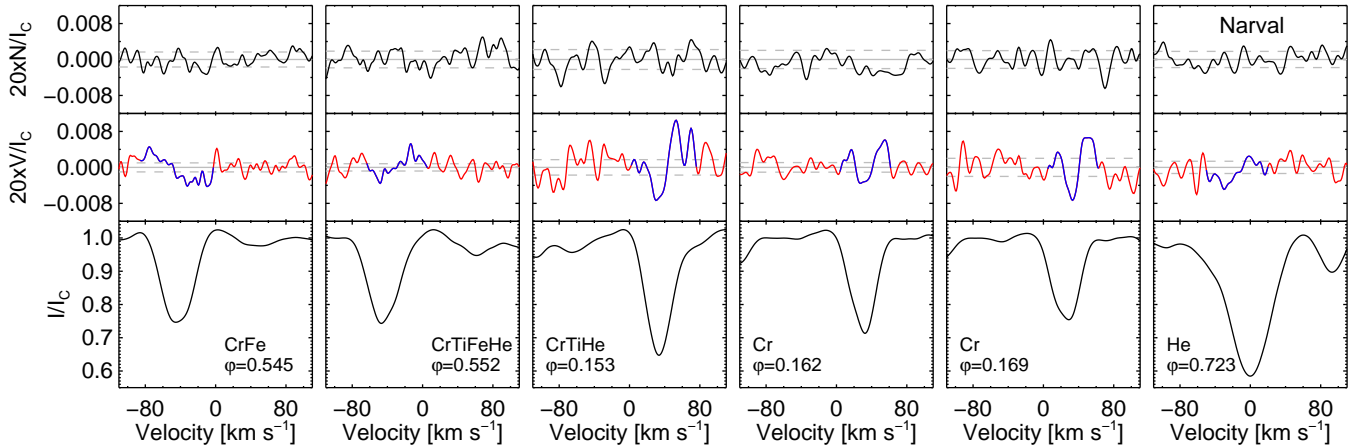}
\caption{
 LSD Stokes~$I$ (bottom), Stokes~$V$ (middle), and diagnostic null (N) profiles (top) calculated for the high-resolution
HARPS\-pol observations of $\upsilon$\,Sgr at five orbital phases using different line masks. Weak Zeeman features are 
highlighted by the blue color. The last plot on the right side
shows Narval observations obtained at the orbital phase $\varphi$=0.723.
The horizontal dashed lines indicate the $\pm1\sigma$ ranges.
}
\label{fig:var_MF}
\end{figure*}

\subsection{Magnetic field measurements}
\label{sect:magn}

\begin{table}
\centering
\caption{
For each orbital phase $\varphi$,
the measured longitudinal magnetic field,
the average Land\'e factor used in the LSD measurements,
the line mask employed,
and the false alarm probability (FAP) values are presented.
} 
\label{tab:log_magf}
\begin{tabular}{ccr@{$\pm$}llcl}
\hline
BJD &
$\varphi$ & 
\multicolumn{2}{c}{$\left<B_{\rm z}\right>$} &
\multicolumn{1}{c}{$\bar{g}_{\rm eff}$} &
Line &
\multicolumn{1}{c}{FAP} \\
2457000+ &
 & 
\multicolumn{2}{c}{(G)} &
 &
mask &
\\
\hline
093.92014 & 0.545 &    22 & 3 & 1.26 & CrFe     & 3.5$\times10^{-5}$\\
094.88218 & 0.552 & $-$10 & 2 & 1.10 & CrTiFeHe & 0.3          \\
177.70278 & 0.153 & $-$40 & 6 & 1.06 & CrTiHe   & 5.1$\times10^{-5}$ \\
178.91667 & 0.162 & $-$19 & 3 & 1.18 & Cr       & 5.4$\times10^{-6}$ \\
179.94422 & 0.169 & $-$25 & 7 & 1.18 & Cr       & 4.9$\times10^{-6}$             \\
\hline
\end{tabular}
\end{table}

The fact that the LSD Stokes~$I$ profiles calculated for the primary component
using different line masks show line profile shapes with different variability character, probably indicating
the presence of chemical patches, needs certainly to be  
taken into account in the analysis of the polarimetric spectra.
Therefore, to search for the presence of typical Zeeman features in the LSD Stokes~$V$ spectra that would indicate the 
presence of a longitudinal magnetic field, we employed in our measurements individual line masks belonging to 
He, Ti, Cr, and Fe and their combinations.
The wide hydrogen lines usually suffer from imperfect continuum normalization and were thus not used.
We present the calculated LSD Stokes~$I$, Stokes~$V$, and diagnostic null ($N$) spectra
for the different line masks in Fig.~\ref{fig:var_MF}.
Null spectra are 
obtained by combining the circularly  polarised spectra in such a way that the
polarisation cancels out, allowing us to verify that no spurious signals are present in the data.
Notably, small Zeeman features are clearly 
visible in the LSD Stokes~$V$ spectra in the observations acquired at four orbital phases.
The measured magnetic fields strengths, the average effective Land\'e factors, the line masks employed, and the False Alarm
Probability values (FAPs) are listed in 
Table~\ref{tab:log_magf}.
FAPs are commonly considered to classify the magnetic field detection in the LSD technique:
${\rm FAP} < 10^{-5}$ is assumed as a definite detection, $10^{-5} < {\rm FAP} < 10^{-3}$ as 
a marginal detection, and ${\rm FAP} > 10^{-3}$ as a non-detection \citep{Donati1992}.
As shown in Table~\ref{tab:log_magf}, we obtain two definite detections at the orbital phases 0.162 and 0.169
with $\bz=-19\pm3$\,G and $\bz=-25\pm7$\,G, correspondingly, and
two marginal detections at the orbital phases 0.545 with $\bz=+22\pm3$\,G, and 0.153 with $\bz=-40\pm6$\,G.

While only a marginal detection, $\bz=22\pm3$\,G, was achieved in the orbital phase 0.545
close to quadrature (see Fig.~1 in \citealt{Koubsky2006}), the weak magnetic field measured
on two consecutive nights in 2015 June in the orbital phases between 0.16 and 0.17 close to the superior 
conjunction is definitely
present. The observed small changes in the field strength are likely caused by the slightly
changing aspect of the overall magnetic field geometry of the star during its rotation
cycle.
We also downloaded the publically available Narval spectra used by \citet{Silvester2009} to search for the
presence of a magnetic field. The last plot on the right side of Fig.~\ref{fig:var_MF} shows
the calculated LSD Stokes~$I$, Stokes~$V$, and diagnostic null ($N$) spectra
for the helium mask using the observation with the highest $S/N$ of 1786 at orbital phase 0.723. A low Zeeman feature with
$\bz=-9\pm2$\,G is still detectable in this plot but ${\rm FAP}=0.124$ indicates non-detection. Clearly, much higher S/N should
be achieved in the spectra to be able to measure such weak magnetic fields. Notably, the higher resolution of 
HARPS\-pol, in comparison to that of Narval, will permit to collect in future observations more photons for a 
given wavelength bin, and, consequently, to achieve a higher $S/N$.

It is of interest that according to \citet{Koubsky2006}, in the orbital cycles when
a blue-shifted H$\alpha$ absorption line is observed, it attains its maximum strength
around the same conjunction phase when the secondary is in front of the primary and
virtually disappears at the other conjunction.
This absorption probably originates from a stream flowing between the two stars.
Since the line was observed on some orbital cycles while missing in others, it is quite
possible that the mass transfer between the two components is governed by the magnetic field
of the primary and the absence of the blue-shifted H$\alpha$ absorption line in
some orbital cycles could be explained by the difference between the lengths of the orbital
and primary rotation periods.

The chemical abundance of the primary in the system 
$\upsilon$\,Sgr was previously reported as peculiar, but, 
as far as we know, no study of the variability of spectral 
lines belonging to different elements was carried out in the past.
The detected variability of the 
LSD Stokes~$I$ profiles calculated for different elements is probably due to a surface inhomogeneous distribution of
the chemical elements similar to that detected in magnetic Ap and Bp stars. In line with this suggestion, a 
number of typical 
chemically peculiar Ap and Bp stars with surface chemical spots were previously reported to host weak longitudinal
magnetic fields 
of the order of just tens of Gauss (e.g.\ \citealt{Donati1990,Donati2006,Kochukhov2018}).
The fact that magnetic field detections are achieved for certain combinations of line masks, indicates that it is possible 
that the chemical element patches are intimately related to the magnetic field geometry of $\upsilon$\,Sgr.
Indeed, previous numerous studies of magnetic Ap and Bp stars have revealed
a symmetry between the geometry of the 
magnetic field and the surface chemical element distribution, i.e. the lines of different elements with different abundance 
distributions across the stellar surface sample the magnetic field in different ways (e.g.\ \citealt{Hubrig2017}). Thus,
the structure of the magnetic fields in Ap and Bp stars can be studied by measurements of the magnetic
field strength using line masks for individual elements separately.

\section{Discussion}
\label{sec:meas}

For the first time, high-resolution spectropolarimetric observations are discussed for the hydrogen-deficient stripped primary 
in the system $\upsilon$\,Sgr, belonging to 
the rare class of evolved binaries that are in a mass transfer phase where the primary has
ended the core helium-burning phase. 
\citet{Schoenberner1983} suggested that the initial  mass of the primary in $\upsilon$\,Sgr must have been 
between 5.6 and 14$\,M_{\odot}$, corresponding to a spectral type in the range B3--B0.5 on the main sequence,
with a radius of the order 4--6\,$R_{\odot}$ \citep{PecautMamajek2013}.
Assuming magnetic flux conservation and 
the stellar radius of 24\,$R_{\odot}$ determined by \citet{Hutter2021},
we can speculate that
the primary possessed on the main sequence a magnetic field of the order of 400--900\,G, which is typical
for magnetic early-type Bp stars.
The system $\upsilon$\,Sgr could therefore have originally been a binary
with a magnetic Bp star that conserved the magnetic flux during the subsequent evolution.

On the other hand, according to the merger scenario, magnetic fields may be 
generated by strong binary interaction, i.e.\ in a stellar merging of two lower-mass stars or protostars,
in the course of mass transfer,
or during common envelope evolution (e.g., \citealt{Tout2008,Ferrario2009,Wickramasinghe2014}).
According to the evolutionary scenario suggested by
\citet{Schoenberner1983}, $\upsilon$\,Sgr
cannot be the result of a single mass exchange.
If $\upsilon$\,Sgr underwent common envelope evolution, this could be responsible for the observed magnetic field 
in the primary of $\upsilon$\,Sgr and it would be responsible for the observed stripping of the envelope.
Since stripped-envelope stars were suggested to be possible progenitors of various types of supernovae including
core-collapse supernovae, more insight into the
uncertain physical processes of binary interaction, both from theoretical and observational sides, would be worthwhile.

Admittedly, the measured longitudinal magnetic field is rather weak. On the other hand, since our observations were obtained at 
five epochs only and the rotation period of the primary is not known, the acquired material is not sufficient to fully 
characterize the geometry and strength of the 
magnetic field. It would be worthwhile to obtain new spectropolarimetric data to learn more
about the structure of the magnetic field in this interacting binary. 

While our observations of the system $\upsilon$\,Sgr mainly targeted
the detection of the magnetic field, it is necessary to understand the three-dimensional
structure of this system with respect to the magnetic field geometry, the orbit, the disk orientation, and the accretion flow.
Importantly, the structure of the circumstellar matter
can only be studied by monitoring $\upsilon$\,Sgr over a significant number of nights.
A future study of the three-dimensional structure of $\upsilon$\,Sgr needs to be based on the intensity variability
of different spectral lines and the measurements of both the radial velocity shifts and the strength of the longitudinal
magnetic field.
It will provide crucial information necessary to test predictions of existing theories on the evolution of massive stars
and allow to gain deeper insights into the role of magnetic fields in binary systems in the context of supernova progenitors.

\section*{Acknowledgements}
Based on observations made with ESO telescopes at the La Silla
Paranal observatory under programme ID 191.D-0255.
We would like to thank the anonymous referee for constructive comments and 
R.~Jayaraman for fruitful discussions.

\section*{Data Availability}

The HARPS spectropolarimetric observations underlying this article can be obtained
from the ESO archive.

\label{lastpage}
\end{document}